\documentclass[twocolumn]{jpsj3}

\usepackage{color}
\usepackage{bm}

\title{%
Gauge-Invariant Cutoff for
Dirac Electron Systems with a Vector Potential
}

\author{%
Yositake Takane
}

\inst{%
Department of Quantum Matter, Graduate School of Advanced Sciences of Matter,\\
Hiroshima University, Higashihiroshima, Hiroshima 739-8530, Japan
}

\recdate{ \hspace{50mm} }

\abst{%
The continuum Dirac model with an unbounded energy spectrum is widely used
to describe low-energy states in various electron systems,
such as graphene, topological insulators, and Weyl semimetals.
However, if it is applied to analyze the electromagnetic response of electrons
to a vector potential,
we often find an unphysical result that breaks gauge invariance.
This is an artifact caused by an energy or wavenumber cutoff,
which is used to avoid divergence of the response.
Here, we propose a modified energy cutoff procedure that preserves
the gauge invariance.
We use this procedure to calculate the response functions
in a two-dimensional massless Dirac electron system.
It is shown that the resulting functions properly describe
the electromagnetic response in a gauge-invariant manner.
}

\begin{document}
\sloppy
\maketitle

\section{Introduction}

In monolayer graphene,~\cite{novoselov1,novoselov2,castro_neto}
electron states near the band touching point
are described by a massless Dirac model, or equivalently, a Weyl model;
thus, they are called Dirac electrons.
Such electron states have been shown to appear in various systems such as
topological insulators,~\cite{fu,moore,roy,ARPES1,ARPES2,ARPES3}
Weyl semimetals,~\cite{shindou,murakami,huang1,xu1}
and $\alpha\rm{\mathchar`-}(BEDT{\mathchar`-}
TTF)_{2}I_{3}$.~\cite{katayama,kobayashi}
These are referred to as Dirac electron systems.
The study on a particular Dirac electron system is traced back to
that on bismuth~\cite{wolff,fukuyama} if the massive case is included.

The electromagnetic response of Dirac electrons has been actively studied
from various aspects.
In the case of graphene, the studies on it have been extended to
orbital magnetism,~\cite{mcclure1,sharma,safran,koshino1}
the screening effect,~\cite{gorbar,ando1,wunsch,hwang}
dynamical conductivity,~\cite{ando2,gusynin,koshino2} and so on.
Here, we focus on a fundamental difficulty that arises in the analysis
based on a continuum Dirac model possessing an unbounded energy spectrum.
When we analyze the response to a vector potential,
the physical quantity under consideration diverges in some cases
owing to the unbounded spectrum.
To avoid such divergence,
we usually introduce a cutoff in energy or wavenumber space.
This cutoff gives rise to an unphysical result that breaks the gauge invariance
with respect to a vector potential $\mib{A}$ and a scalar potential $\phi$.
A typical example is that if a charge current density $\mib{j}$ is calculated
in response to $\mib{A}$, we erroneously find that $\mib{j}$ becomes finite
even when $\mib{A}$ is constant.~\cite{principi,polini,katsnelson}
Needless to say, a constant vector potential induces no effect on
a physical system owing to the gauge invariance.
Similar difficulties arise in the analysis of, for example,
the superfluid density of Dirac electrons
in the superconducting state~\cite{kopnin,mizoguchi}
and the chiral magnetic effect in a Weyl semimetal.~\cite{takane1,takane2}
Although the insufficiency of an energy or wavenumber cutoff
has been recognized,~\cite{principi,polini}
little attempt has been made to overcome this difficulty.~\cite{juricic}

In this paper, we propose a modified energy cutoff procedure
for general Dirac electron systems
to improve the description of their electromagnetic response.
Its original form is briefly reported in Ref.~\citen{takane2}
in an incomplete manner.
The modified energy cutoff preserves the gauge invariance
and removes a difficulty
that arises in the analysis of the response to a vector potential.
We use the modified energy cutoff procedure to calculate
the charge and current densities induced by a vector potential
in a two-dimensional (2D) massless Dirac electron system.
We show that it enables us to describe the electromagnetic response
in a gauge-invariant manner.

In the next section, we present a 2D massless Dirac model
with a single valley, and derive the response functions
for vector and scalar potentials by using an ordinary cutoff.
The resulting response functions break the gauge invariance
as well as the charge conservation relation.
In Sect.~3, we propose a modified energy cutoff procedure
and roughly show how it works.
In Sect.~4, we derive the response functions by applying
the modified energy cutoff procedure.
The resulting response functions preserve the gauge invariance
and satisfy the charge conservation relation.
In Sect.~5, the modified energy cutoff procedure is justified
in an accurate manner.
The last section is devoted to a short summary.
We set $\hbar = k_{\rm B} = 1$ throughout this paper.

\section{Model, Formulation, and Known Results}

We introduce the 2D massless Dirac model with a single Dirac cone
centered at $\mib{k}=(0,0)$:~\cite{mcclure2,slonczewski,ajiki}
\begin{align}
        \label{eq:Hamiltonian}
   H
 & = \int d^{2}r
     \psi^{\dagger}(\mib{r})
     \Bigl[ v\left(\sigma_{x}\hat{k}_{x}+\sigma_{y}\hat{k}_{y}\right)
            - \mu
     \Bigr]
     \psi(\mib{r}) ,
\end{align}
where $\psi(\mib{r})$ represents the spinor field describing
Dirac electrons, and $v$ and $\mu$ respectively
denote the velocity and chemical potential.
Here, $\sigma_{x}$ and $\sigma_{y}$ are the $x$- and $y$-components
of the Pauli matrix,
and $\hat{k}_{x}=-i\partial_{x}$ and $\hat{k}_{y}=-i\partial_{y}$.
The eigenvalue of energy is determined as
\begin{align}
  E_{\eta}(\mib{k}) = \eta v|\mib{k}|-\mu ,
\end{align}
where $\eta = +$ for the conduction band and $\eta = -$ for the valence band.
The perturbations due to the vector potential $\mib{A}$
and the scalar potential $\phi$ are respectively expressed as
\begin{align}
   H_{A}
  & = - \int d^{2}r\, \mib{j}(\mib{r})\cdot \mib{A}(\mib{r}) ,
      \\
   H_{\phi}
  & = \int d^{2}r\, \rho(\mib{r})\phi(\mib{r}) .
\end{align}
The charge current density $\mib{j}=(j_{x},j_{y})$
and the charge density $\rho$ are expressed as
\begin{align}
       \label{eq:def-current}
  \mib{j} & = - ev \psi^{\dagger}(\mib{r})
                   \left(\sigma_{x},\sigma_{y}\right)\psi(\mib{r}) ,
      \\
       \label{eq:def-charge}
  \rho & = - e \psi^{\dagger}(\mib{r})\psi(\mib{r}) .
\end{align}

We consider the current and charge densities induced by
the vector potential in the $x$-direction
\begin{align}
 \mib{A} = (A_{x}(\mib{q},\omega),0)e^{i\mib{q}\cdot\mib{r}-i\omega t}
\end{align}
or the scalar potential
\begin{align}
 \phi = \phi(\mib{q},\omega)e^{i\mib{q}\cdot\mib{r}-i\omega t} .
\end{align}
Within linear response theory, the average current and charge densities
are expressed by the response functions $\chi_{\alpha \Gamma}$
with $\alpha = j$, $\rho$ and
$\Gamma = A$, $\phi$.~\cite{wunsch,principi,sabio,stauber}
They are defined so that
the average current and charge densities are expressed as
\begin{align}
  \langle j_{x}(\mib{q},\omega) \rangle_{A}
 & = -e^{2}v^{2}\chi_{j A}(\mib{q},\omega)A_{x}(\mib{q},\omega) ,
      \\
  \langle \rho(\mib{q},\omega) \rangle_{A}
 & = -e^{2}v\chi_{\rho A}(\mib{q},\omega)
      A_{x}(\mib{q},\omega) ,
      \\
  \langle j_{x}(\mib{q},\omega) \rangle_{\phi}
 & = -e^{2}v\chi_{j \phi}(\mib{q},\omega)\phi(\mib{q},\omega) ,
      \\
  \langle \rho(\mib{q},\omega) \rangle_{\phi}
 & = -e^{2}\chi_{\rho \phi}(\mib{q},\omega)
      \phi(\mib{q},\omega) .
\end{align}
The response functions are obtained by performing the analytic continuation
of $i\nu \to \omega+i\delta$ from their Matsubara representation,
\begin{align}
           \label{eq:def-res-jj}
    \Pi_{j A}(\mib{q},i\nu)
  &  = \int\frac{d^{2}k}{(2\pi)^{2}}T\sum_{\epsilon}
          \nonumber \\
  &    \hspace{-2mm} \times
       {\rm tr} \left\{\sigma_{x}G(\mib{k}+\mib{q},i\epsilon+i\nu)
                       \sigma_{x}G(\mib{k},i\epsilon)\right\} ,
       \\
           \label{eq:def-res-rj}
    \Pi_{\rho A}(\mib{q},i\nu)
  &  = \int\frac{d^{2}k}{(2\pi)^{2}}T\sum_{\epsilon}
          \nonumber \\
  &    \hspace{-2mm} \times
       {\rm tr} \left\{G(\mib{k}+\mib{q},i\epsilon+i\nu)
                       \sigma_{x}G(\mib{k},i\epsilon)\right\} ,
       \\
           \label{eq:def-res-jr}
    \Pi_{j \phi}(\mib{q},i\nu)
  &  = -\int\frac{d^{2}k}{(2\pi)^{2}}T\sum_{\epsilon}
          \nonumber \\
  &    \hspace{-2mm} \times
       {\rm tr} \left\{\sigma_{x}G(\mib{k}+\mib{q},i\epsilon+i\nu)
                       G(\mib{k},i\epsilon)\right\} ,
       \\
           \label{eq:def-res-rr}
    \Pi_{\rho \phi}(\mib{q},i\nu)
  &  = -\int\frac{d^{2}k}{(2\pi)^{2}}T\sum_{\epsilon}
          \nonumber \\
  &    \hspace{-2mm} \times
       {\rm tr} \left\{G(\mib{k}+\mib{q},i\epsilon+i\nu)
                       G(\mib{k},i\epsilon)\right\} ,
\end{align}
where $T$ is the temperature.
Here, the thermal Green's function is given by
\begin{align}
  G(\mib{k},i\epsilon)
  = \frac{1}{2}\sum_{\eta = \pm}
    \frac{1+ \eta\left(\sigma_{x}\cos\varphi_{\mib{k}}
                       + \sigma_{y}\sin\varphi_{\mib{k}}\right)}
         {i\epsilon - E_{\eta}(\mib{k})} .
\end{align}
where
\begin{align}
  \cos\varphi_{\mib{k}} = \frac{k_{x}}{|\mib{k}|} ,
  \hspace{5mm}
  \sin\varphi_{\mib{k}} = \frac{k_{y}}{|\mib{k}|} .
\end{align}
After performing the Matsubara summation, we find
\begin{align}
      \label{eq:Pi-jA}
  \Pi_{j A}(\mib{q},i\nu)
  & = \int\frac{d^{2}k}{(2\pi)^{2}}
      \sum_{\eta,\eta'=\pm}
      \frac{1+\eta\eta'\cos\left(\varphi_{\mib{k}}+\varphi_{\mib{k}+\mib{q}}
                           \right)}{2}
          \nonumber \\
  & \hspace{3mm} \times
      \frac{f_{\rm FD}(E_{\eta}(\mib{k}))
            - f_{\rm FD}(E_{\eta'}(\mib{k}+\mib{q}))}
           {i\nu + E_{\eta}(\mib{k}) - E_{\eta'}(\mib{k}+\mib{q})} ,
       \\
      \label{eq:Pi-rA}
  \Pi_{\rho A}(\mib{q},i\nu)
  & = \int\frac{d^{2}k}{(2\pi)^{2}}
      \sum_{\eta,\eta'=\pm}
      \frac{\eta\cos\varphi_{\mib{k}} + \eta'\cos\varphi_{\mib{k}+\mib{q}}}{2}
          \nonumber \\
  & \hspace{3mm} \times
      \frac{f_{\rm FD}(E_{\eta}(\mib{k}))
            - f_{\rm FD}(E_{\eta'}(\mib{k}+\mib{q}))}
           {i\nu + E_{\eta}(\mib{k}) - E_{\eta'}(\mib{k}+\mib{q})} ,
       \\
      \label{eq:Pi-rp}
  \Pi_{\rho \phi}(\mib{q},i\nu)
  & = -\int\frac{d^{2}k}{(2\pi)^{2}}
      \sum_{\eta,\eta'=\pm}
      \frac{1+\eta\eta'\cos\left(\varphi_{\mib{k}}-\varphi_{\mib{k}+\mib{q}}
                           \right)}{2}
          \nonumber \\
  & \hspace{3mm} \times
      \frac{f_{\rm FD}(E_{\eta}(\mib{k}))
            - f_{\rm FD}(E_{\eta'}(\mib{k}+\mib{q}))}
           {i\nu + E_{\eta}(\mib{k}) - E_{\eta'}(\mib{k}+\mib{q})} ,
       \\
  \Pi_{j \phi}(\mib{q},i\nu)
  & = - \Pi_{\rho A}(\mib{q},i\nu) ,
\end{align}
where $f_{\rm FD}(E)$ represents the Fermi--Dirac function.

For simplicity, we hereafter focus on the case of $\mu = 0$ at $T = 0$.
Equations~(\ref{eq:Pi-jA})--(\ref{eq:Pi-rp}) indicate that
$\Pi_{\alpha \Gamma}$ generally consists of the interband contribution
arising from the terms with $\eta \neq \eta'$
and the intraband contribution arising from those with $\eta = \eta'$.
In this case, only the interband terms contribute to the response functions,
reflecting the fact that $f_{\rm FD}(E_{+}(\mib{k})) = 0$
and $f_{\rm FD}(E_{-}(\mib{k})) = 1$ for any $\mib{k}$.
Let us consider the response to
$\mib{A} = (A_{x}(\mib{q},\omega),0)e^{i\mib{q}\cdot\mib{r}-i\omega t}$.
For this vector potential, we need to separately treat the transverse case
with $\mib{q} = (0,q)$ and the longitudinal case with $\mib{q} = (q,0)$.
Since the response functions describing $j_{x}$ diverge
without a regularization,~\cite{principi}
we employ an ordinary energy cutoff at $E = -\varepsilon_{M}$
that restricts the integration over $\mib{k}$ by the condition of
$|\mib{k}| < k_{M}$, where $k_{M} = \varepsilon_{M}/v$.
The response function describing $\rho$ converges without a regularization.
The response functions are given as
\begin{align}
     \label{eq:chi-jA-t_oc}
  \chi_{j A}^{t,{\rm oc}}(q\hat{\mib{y}},\omega)
  & = \frac{\sqrt{(vq)^{2}-(\omega+i\delta)^{2}}}{16v^{2}}
      - \frac{\varepsilon_{M}}{4\pi v^{2}} ,
      \\
     \label{eq:chi-jA-l_oc}
  \chi_{j A}^{l,{\rm oc}}(q\hat{\mib{x}},\omega)
  & = - \frac{\omega^{2}}{16v^{2}\sqrt{(vq)^{2}-(\omega+i\delta)^{2}}}
      - \frac{\varepsilon_{M}}{4\pi v^{2}} ,
      \\
  \chi_{\rho A}^{l}(q\hat{\mib{x}},\omega)
  & = - \frac{\omega (vq)}{16v^{2}\sqrt{(vq)^{2}-(\omega+i\delta)^{2}}} ,
\end{align}
where $t$ and $l$ respectively represent the transverse and longitudinal cases,
and ${\rm oc}$ indicates that the corresponding result
is obtained by using the ordinary cutoff.
Note that $\chi_{\rho A}^{t}(q\hat{\mib{y}},\omega) = 0$
as a transverse vector potential cannot induce a charge density.
Equations~(\ref{eq:chi-jA-t_oc}) and (\ref{eq:chi-jA-l_oc})
have been given in Ref.~\citen{principi}.
Let us next consider the response to
$\phi = \phi(\mib{q},\omega) e^{i\mib{q}\cdot\mib{r}-i\omega t}$.
The response functions converge without a regularization,~\cite{ando1,wunsch}
resulting in
\begin{align}
     \label{eq:chi-jp}
  \chi_{j \phi}(q,\omega)
  & = \frac{\omega (vq)}{16v^{2}\sqrt{(vq)^{2}-(\omega+i\delta)^{2}}} ,
      \\
     \label{eq:chi-rp}
  \chi_{\rho \phi}(q,\omega)
  & = \frac{(vq)^{2}}{16v^{2}\sqrt{(vq)^{2}-(\omega+i\delta)^{2}}} .
\end{align}
Equation~(\ref{eq:chi-rp}) has been given in Ref.~\citen{wunsch}.
For $(vq)^{2} < \omega^{2}$, the square roots in the above expressions
should be read as $\sqrt{(vq)^{2}-(\omega+i\delta)^{2}}
= -i\,{\rm sign}(\omega)\sqrt{\omega^{2}-(vq)^{2}}$.

It is easy to observe that these response functions break the gauge invariance
with respect to $\mib{A}$ and $\phi$.~\cite{principi,polini}
Owing to the gauge invariance, a static transverse vector potential can induce
no electromagnetic response in the limit of $q \to 0$,
while a static longitudinal vector potential
cannot induce a response for any $q$.
Contrary to this well-known fact, the term with $\varepsilon_{M}$
in $\chi_{j A}^{t,{\rm oc}}$ and $\chi_{j A}^{l,{\rm oc}}$
induces a finite charge current even when $\mib{A}$ is constant.
Furthermore, the gauge invariance ensures that $\langle j_{x} \rangle_{A}$
induced by $A_{x}(q\hat{\mib{x}},\omega)$ must be identical to
$\langle j_{x} \rangle_{\phi}$ induced by
$\phi(q\hat{\mib{x}},\omega) \equiv (-\omega/q)A_{x}(q\hat{\mib{x}},\omega)$.
However, they apparently differ from each other.
Indeed, we find
\begin{align}
  \langle j_{x} \rangle_{A} - \langle j_{x} \rangle_{\phi}
  = ev \frac{\varepsilon_{M}}{4\pi v}
       eA_{x}(q\hat{\mib{x}},\omega) .
\end{align}
Note that the response to a scalar potential satisfies
the charge conservation relation
\begin{align}
  -\omega\langle \rho \rangle_{\phi} + q\langle j_{x} \rangle_{\phi} = 0 ,
\end{align}
whereas the response to a vector potential breaks it as
\begin{align}
  -\omega\langle \rho \rangle_{A} + q\langle j_{x} \rangle_{A}
  = e vq\frac{\varepsilon_{M}}{4\pi v} eA_{x}(q\hat{\mib{x}},\omega) .
\end{align}
The above argument indicates that,
although the response to a scalar potential is appropriate,
we need to reconsider the response to a vector potential.
It has been pointed out that this difficulty is caused by the ordinary cutoff,
which breaks the gauge invariance.~\cite{principi,polini}

\section{Gauge-Invariant Energy Cutoff}

To overcome the difficulty associated with $\mib{A}$,
we propose a modified energy cutoff that preserves the gauge invariance.
This cutoff is implemented by two steps.
Firstly, we replace the Fermi--Dirac function $f_{\rm FD}(E)$
in the expression for a response function with the modified distribution
function $\tilde{f}_{\rm FD}(E)$ defined by
\begin{align}
  \tilde{f}_{\rm FD}(E) = f_{\rm FD}(E)\theta(E+\varepsilon_{M}) ,
\end{align}
where $\theta(E)$ is the Heaviside step function.
Secondly, we calculate the correction induced by this replacement
in the zero-frequency limit of $\omega \to 0$.
By adding the resulting correction to the main contribution given in Sect.~2,
we obtain the final result [see Eq.~(\ref{eq:def-mc-t}) as an example].

The replacement of $f_{\rm FD}(E)$ with $\tilde{f}_{\rm FD}(E)$
directly results in the exclusion of the electron states
with energy $E$ smaller than $-\varepsilon_{M}$.
This is not equivalent to the restriction of $|\mib{k}| < k_{M}$
in the ordinary cutoff, as shown below.
Indeed, it gives a new correction by which the gauge invariance is preserved.
The zero-frequency limit is taken to pick up
only the relevant correction in a selective manner.
In other words, a spurious contribution is included without taking the limit.
Accurate justification of this procedure is given in Sect.~5.

Now, we briefly point out an essential difference between
the ordinary cutoff and the modified one proposed here.
If the ordinary cutoff at $E = -\varepsilon_{M}$
is applied to the calculation of a response function, only an initial state
is restricted to satisfy the condition of $-\varepsilon_{M} < E$.
In other words, the restriction is not imposed on an intermediate state.
By using the modified energy cutoff, we can thoroughly impose the restriction
of $-\varepsilon_{M} < E$ both on initial and intermediate states.
We here elucidate the importance of this difference
by applying these two procedures to a simple problem.

Let us consider the variation in the total energy of electron states induced by
a static vector potential $\mib{A} = (A_{x},0)e^{i\mib{q}\cdot\mib{x}}$
in the limit of $|\mib{q}| \to 0$.
Obviously, as the resulting vector potential is constant,
it never alters the total energy owing to the gauge invariance.
We again focus on the case of $\mu = 0$ at $T = 0$,
and calculate the variation $\delta U$
within a second-order perturbation theory with respect to $\mib{A}$.
If the ordinary cutoff is applied,
the variation arises only from interband processes and is expressed as
\begin{align}
 \delta U_{\rm inter}
 & = \left(veA_{x}\right)^{2}\int\frac{d^{2}k}{(2\pi)^{2}}
     \sin^{2}\left(\frac{\varphi_{\mib{k}}+\varphi_{\mib{k}+\mib{q}}}{2}\right)
        \nonumber \\
 & \hspace{5mm} \times
     \frac{f_{\rm FD}(E_{-}(\mib{k}))
           \left[1-f_{\rm FD}(E_{+}(\mib{k}+\mib{q}))\right]}
          {E_{-}(\mib{k}) - E_{+}(\mib{k}+\mib{q})} ,
\end{align}
where the integration over $\mib{k}$ is restricted by $k < k_{M}$
with $k = |\mib{k}|$.
In the limit of $|\mib{q}| \to 0$, we find
\begin{align}
      \label{eq:inter-U}
  \delta U_{\rm inter}
  = - \left(eA_{x}\right)^{2}\frac{\varepsilon_{M}}{8\pi} ,
\end{align}
which disagrees with the correct result, $\delta U = 0$,
expected from the gauge invariance.
This clearly indicates that
the ordinary cutoff breaks the gauge invariance.

We show that the correct result is obtained
if the modified energy cutoff is applied.~\cite{comment0}
The variation arises from not only interband processes
but also intraband processes.
The former contribution is identical to that given in Eq.~(\ref{eq:inter-U}).
The latter contribution is expressed as
\begin{align}
 \delta U_{\rm intra}
 & = \left(veA_{x}\right)^{2}\int\frac{d^{2}k}{(2\pi)^{2}}
     \cos^{2}\left(\frac{\varphi_{\mib{k}}+\varphi_{\mib{k}+\mib{q}}}{2}\right)
        \nonumber \\
 & \hspace{5mm} \times
     \frac{\tilde{f}_{\rm FD}(E_{-}(\mib{k}))
           \left[1-\tilde{f}_{\rm FD}(E_{-}(\mib{k}+\mib{q}))\right]}
          {E_{-}(\mib{k}) - E_{-}(\mib{k}+\mib{q})} .
\end{align}
As a direct consequence of the restriction on the intermediate state
with $E_{-}(\mib{k}+\mib{q})$, this gives a nonnegligible contribution
when $E_{-}(\mib{k}+\mib{q}) < -\varepsilon_{M} < E_{-}(\mib{k})$.
Approximating the fractional factor as
\begin{align}
  & \frac{\tilde{f}_{\rm FD}(E_{-}(\mib{k}))
          \left[1-\tilde{f}_{\rm FD}(E_{-}(\mib{k}+\mib{q}))\right]}
         {E_{-}(\mib{k}) - E_{-}(\mib{k}+\mib{q})}
       \nonumber \\
  & \hspace{10mm}
    = \theta(E_{-}(\mib{k})+\varepsilon_{M})
      \delta(E_{-}(\mib{k})+\varepsilon_{M}) ,
\end{align}
we find
\begin{align}
      \label{eq:intra-U}
  \delta U_{\rm intra}
  = \left(eA_{x}\right)^{2}\frac{\varepsilon_{M}}{8\pi} ,
\end{align}
which exactly cancels out $\delta U_{\rm inter}$.
That is, the modified energy cutoff gives the correct result,
\begin{align}
  \delta U = \delta U_{\rm inter} + \delta U_{\rm intra} = 0 .
\end{align}
This argument suggests that
the modified energy cutoff is more suitable than the ordinary one
in describing the response to a vector potential in Dirac electron systems.

The insufficiency of the ordinary cutoff is clearly explained
from the fact that a constant vector potential $\mib{A} = (A_{x},0)$
only shifts the Dirac point from $\mib{k} = (0,0)$ to $(-eA_{x},0)$.
In the absence of $\mib{A}$, the energy cutoff at $E = -\varepsilon_{M}$ is
equivalent to restricting the integration over $\mib{k}$
by the condition of $k < k_{M}$.
In the presence of $\mib{A}$, the energy cutoff is correctly carried out
by modifying the condition as
$\sqrt{(k_{x}+eA_{x})^{2}+k_{y}^{2}} < k_{M}$.
The ordinary cutoff takes no account of such a modification;
thus, it breaks the gauge invariance.
Indeed, we can show that $\delta U_{\rm inter}$ is identical to the variation
in the total energy under the ordinary cutoff:
\begin{align}
  \delta U_{\rm oc}
  = \int\frac{d^{2}k}{(2\pi)^{2}}
    \left( E_{-}(\mib{k}+e\mib{A}) - E_{-}(\mib{k}) \right) ,
\end{align}
where the integration over $\mib{k}$ is restricted by $k < k_{M}$.
In the modified energy cutoff, the modification of the condition
is implicitly taken into account through $\tilde{f}_{\rm FD}(E)$.

\section{Derivation of Response Functions}

By using the modified energy cutoff proposed in Sect.~3, we derive
the response functions for $\mib{A}$ in the case of $\mu = 0$ at $T = 0$.
The response functions $\chi_{j A}^{t,{\rm oc}}(q\hat{\mib{y}},\omega)$
and $\chi_{j A}^{l,{\rm oc}}(q\hat{\mib{x}},\omega)$
given in Sect.~2 are obtained by using the ordinary cutoff
and consist of only the interband contribution
arising from the terms with $\eta \neq \eta'$.
Even though the modified energy cutoff is applied instead of
the ordinary one, the interband contribution does not change.
However, the intraband term with $\eta = \eta' = -$
gives an additional contribution.
Hence, we derive this contribution $\delta\Pi_{j A}(\mib{q},i\nu)$
in the Matsubara representation.
According to the procedure given in Sect.~3, the proper correction is obtained
by taking the zero-frequency limit of $\omega \to 0$
after the analytic continuation of $i\nu \to \omega +i\delta$.
For example, the correction to $\chi_{j A}^{t,{\rm oc}}(q\hat{\mib{y}},\omega)$
is given by
\begin{align}
     \label{eq:def-corre-t}
  \delta\chi_{j A}^{t}(q\hat{\mib{y}},0)
  = \lim_{\omega \to 0}
    \left[ \left. \delta\Pi_{j A}(q\hat{\mib{y}},i\nu)
           \right|_{i\nu \to \omega+i\delta}
    \right] .
\end{align}
The final result is expressed as
\begin{align}
     \label{eq:def-mc-t}
  \chi_{j A}^{t,{\rm mc}}(q\hat{\mib{y}},\omega)
  = \chi_{j A}^{t,{\rm oc}}(q\hat{\mib{y}},\omega)
    + \delta\chi_{j A}^{t}(q\hat{\mib{y}},0) ,
\end{align}
where ${\rm mc}$ indicates that
this is obtained by using the modified energy cutoff.

From Eq.~(\ref{eq:Pi-jA}),
we find that the additional contribution is expressed as
\begin{align}
  \delta\Pi_{j A}(\mib{q},i\nu)
  & = \int\frac{d^{2}k}{(2\pi)^{2}}
      \frac{1+\cos\left(\varphi_{\mib{k}}+\varphi_{\mib{k}+\mib{q}}
                           \right)}{2}
          \nonumber \\
  & \hspace{-3mm} \times
      \frac{\tilde{f}_{\rm FD}(E_{-}(\mib{k}))
            - \tilde{f}_{\rm FD}(E_{-}(\mib{k}+\mib{q}))}
           {i\nu + E_{-}(\mib{k}) - E_{-}(\mib{k}+\mib{q})} .
\end{align}
A nonnegligible contribution arises from the cases of
$E_{-}(\mib{k}+\mib{q}) < -\varepsilon_{M} < E_{-}(\mib{k})$
and $E_{-}(\mib{k}) < -\varepsilon_{M} < E_{-}(\mib{k}+\mib{q})$.
This can be safely calculated by using the following approximation:
\begin{align}
  & \tilde{f}_{\rm FD}(E_{-}(\mib{k}))
       -\tilde{f}_{\rm FD}(E_{-}(\mib{k}+\mib{q}))
       \nonumber \\
  & \hspace{5mm}
    = -\frac{\partial \tilde{f}_{\rm FD}(E_{-}(\mib{k}))}
            {\partial E_{-}(\mib{k})}
       \left(-vq\sin\varphi_{\mib{k}}\right) ,
\end{align}
for a transverse vector potential with $\mib{q} = (0,q)$.
For a longitudinal vector potential with $\mib{q} = (q,0)$,
the factor $-vq\sin\varphi_{\mib{k}}$ in the right-hand side should be
replaced with $-vq\cos\varphi_{\mib{k}}$.
The integration over $k$ yields
\begin{align}
     \label{eq:exp-deltaPi-t}
  \delta\Pi_{j A}(q\hat{\mib{y}},i\nu)
    = \frac{\varepsilon_{M}}{2\pi v^{2}}
      \int_{0}^{2\pi}\frac{d\varphi}{2\pi}
      \cos^{2}\varphi \frac{vq\sin\varphi}{i\nu+vq\sin\varphi}
\end{align}
for a transverse vector potential.
The correction to $\chi_{j A}^{t,{\rm oc}}(q\hat{\mib{y}},\omega)$
is obtained by using Eq.~(\ref{eq:def-corre-t}).

We find that the resulting corrections to $\chi_{j A}^{t,{\rm oc}}$
and $\chi_{j A}^{l,{\rm oc}}$ are equivalent and are given by
\begin{align}
   \delta\chi_{j A}^{t}(q\hat{\mib{y}},0)
   = \delta\chi_{j A}^{l}(q\hat{\mib{x}},0)
   = \frac{\varepsilon_{M}}{4\pi v^{2}} ,
\end{align}
which exactly cancels out the second term of $\chi_{j A}^{t,{\rm oc}}$
and $\chi_{j A}^{l,{\rm oc}}$.
No intraband correction appears in $\chi_{\rho A}^{l}$.
By using the modified energy cutoff, we finally find that
the response functions for a vector potential are given by
\begin{align}
    \chi_{j A}^{t,{\rm mc}}(q\hat{\mib{y}},\omega)
 &  = \frac{\sqrt{(vq)^{2}-(\omega+i\delta)^{2}}}{16v^{2}} ,
      \\
    \chi_{j A}^{l,{\rm mc}}(q\hat{\mib{x}},\omega)
 &  = - \frac{\omega^{2}}{16v^{2}\sqrt{(vq)^{2}-(\omega+i\delta)^{2}}} ,
      \\
  \chi_{\rho A}^{l}(q\hat{\mib{x}},\omega)
  & = - \frac{\omega (vq)}{16v^{2}\sqrt{(vq)^{2}-(\omega+i\delta)^{2}}} .
\end{align}
The electromagnetic response is described by these response functions
together with those given in Eqs.~(\ref{eq:chi-jp}) and (\ref{eq:chi-rp}).

In contrast to the results under the ordinary cutoff,
the response functions $\chi_{j A}^{t,{\rm mc}}$ and
$\chi_{j A}^{l,{\rm mc}}$ together with $\chi_{\rho A}^{l}$
satisfy the conditions required from the gauge invariance.
Indeed, $\chi_{j A}^{t,{\rm mc}}(q\hat{\mib{y}},0) = 0$
in the limit of $q \to 0$ and $\chi_{j A}^{l,{\rm mc}}(q\hat{\mib{x}},0)
= \chi_{\rho A}^{l}(q\hat{\mib{x}},0) = 0$ for any $q$.
In addition, we can show that $\langle j_{x} \rangle_{A}$ induced by
$A_{x}(q\hat{\mib{x}},\omega)$ is identical to $\langle j_{x} \rangle_{\phi}$
induced by
$\phi(q\hat{\mib{x}},\omega) \equiv (-\omega/q)A_{x}(q\hat{\mib{x}},\omega)$.
We can also show that the charge conservation relation holds
in the response to a longitudinal vector potential as
\begin{align}
  -\omega\langle \rho \rangle_{A} + q\langle j_{x} \rangle_{A} = 0 .
\end{align}
The above argument indicates that the gauge invariance is preserved
if we use the modified energy cutoff to calculate
the response functions for $\mib{A}$.

The resulting response functions are equivalent to $\chi_{j A}^{t,{\rm oc}}$
and $\chi_{j A}^{l,{\rm oc}}$ obtained by using the ordinary cutoff
if the term $-\varepsilon_{M}/(4\pi v^{2})$ is simply excluded.
However, note that this exclusion has not been justified in a reliable manner.
Indeed, it was claimed~\cite{principi} that this term is physical
in the limit of $q \to 0$ with $\omega \neq 0$.

\section{Justification of the Modified Energy Cutoff}

In this section, we derive all the response functions
by applying the modified energy cutoff procedure
\textit{without taking the zero-frequency limit of} $\omega \to 0$.
The resulting response functions  $\tilde{\chi}_{\alpha \Gamma}$
consist of two contributions: a relevant contribution
that describes the actual response of Dirac electrons
and an irrelevant contribution
that reflects the effect of artificial excitations due to the cutoff.
We show that the zero-frequency limit allows us to pick up only
the relevant contribution, justifying the procedure given in Sect.~3.

We start with the expression of $\delta\Pi_{j A}$, given in
Eq.~(\ref{eq:exp-deltaPi-t}), for a transverse vector potential.
Performing the integration over $\varphi$, we find
\begin{align}
  \delta\Pi_{j A}(q\hat{\mib{y}},i\nu)
    = \frac{\varepsilon_{M}}{4\pi v^{2}}
      \left[ 1-\frac{2\nu\left(\sqrt{(vq)^{2}+\nu^{2}}-\nu\right)}
                    {(vq)^{2}}
      \right] .
\end{align}
The corresponding correction to $\chi_{j A}^{t,{\rm oc}}$ is obtained
by performing the analytic continuation of $i\nu \to \omega +i\delta$.
The result is written as
\begin{align}
       \label{eq:del-jAt}
 &  \delta\chi_{j A}^{t}(q\hat{\mib{y}},\omega)
        \nonumber \\
 & \hspace{0mm}
    = \frac{\varepsilon_{M}}{4\pi v^{2}}
      \left[ 1+\frac{2i\omega\sqrt{(vq)^{2}-(\omega+i\delta)^{2}}}{(vq)^{2}}
               \Lambda(q,\omega)
      \right] ,
\end{align}
where
\begin{align}
   \Lambda(q,\omega)
   = 1+\frac{i\omega}{\sqrt{(vq)^{2}-(\omega+i\delta)^{2}}} .
\end{align}
Performing calculations similar to this, we find that
the corrections to the other response functions are
\begin{align}
       \label{eq:del-jAl}
  \delta\chi_{j A}^{l}(q\hat{\mib{x}},\omega)
  & = \frac{\varepsilon_{M}}{4\pi v^{2}}
      \left[ 1+\frac{2 \omega^{2}}{(vq)^{2}}\Lambda(q,\omega) \right] ,
    \\
       \label{eq:del-rAl}
  \delta\chi_{\rho A}^{l}(q\hat{\mib{x}},\omega)
  & = \frac{\varepsilon_{M}}{2\pi v^{2}}
      \frac{\omega}{vq}\Lambda(q,\omega) ,
    \\
       \label{eq:del-jp}
  \delta\chi_{j \phi}(q,\omega)
  & = - \frac{\varepsilon_{M}}{2\pi v^{2}}
      \frac{\omega}{vq}\Lambda(q,\omega) ,
    \\
       \label{eq:del-rp}
  \delta\chi_{\rho \phi}(q,\omega)
  & = - \frac{\varepsilon_{M}}{2\pi v^{2}}\Lambda(q,\omega) .
\end{align}

Adding each correction to the corresponding main contribution given in Sect.~2,
we finally find the response functions $\tilde{\chi}_{\alpha \Gamma}$.
The results are
\begin{align}
     \label{eq:tilde-jAt}
    \tilde{\chi}_{j A}^{t}(q\hat{\mib{y}},\omega)
 &  = \frac{\sqrt{(vq)^{2}-(\omega+i\delta)^{2}}}{16v^{2}}
           \nonumber \\
 & \hspace{0mm}
     + \frac{\varepsilon_{M}}{2\pi v^{2}}
       \frac{i\omega\sqrt{(vq)^{2}-(\omega+i\delta)^{2}}}{(vq)^{2}}
       \Lambda(q,\omega) ,
      \\
     \label{eq:tilde-jAl}
    \tilde{\chi}_{j A}^{l}(q\hat{\mib{x}},\omega)
 &  = - \frac{\omega^{2}}{16v^{2}\sqrt{(vq)^{2}-(\omega+i\delta)^{2}}}
           \nonumber \\
 & \hspace{0mm}
     + \frac{\varepsilon_{M}}{2\pi v^{2}}\frac{\omega^{2}}{(vq)^{2}}
        \Lambda(q,\omega) ,
      \\
     \label{eq:tilde-rA}
    \tilde{\chi}_{\rho A}^{l}(q\hat{\mib{x}},\omega)
 &  = - \frac{\omega (vq)}{16v^{2}\sqrt{(vq)^{2}-(\omega+i\delta)^{2}}}
           \nonumber \\
 & \hspace{0mm}
     + \frac{\varepsilon_{M}}{2\pi v^{2}}\frac{\omega}{vq}
        \Lambda(q,\omega) ,
      \\
     \label{eq:tilde-jp}
    \tilde{\chi}_{j \phi}(q,\omega)
 &  = \frac{\omega (vq)}{16v^{2}\sqrt{(vq)^{2}-(\omega+i\delta)^{2}}}
           \nonumber \\
 & \hspace{0mm}
     - \frac{\varepsilon_{M}}{2\pi v^{2}}\frac{\omega}{vq}
        \Lambda(q,\omega) ,
      \\
     \label{eq:tilde-rp}
    \tilde{\chi}_{\rho \phi}(q,\omega)
 &  = \frac{(vq)^{2}}{16v^{2}\sqrt{(vq)^{2}-(\omega+i\delta)^{2}}}
     - \frac{\varepsilon_{M}}{2\pi v^{2}}\Lambda(q,\omega) .
\end{align}
It is easy to see that these response functions preserve the gauge invariance
and satisfy the charge conservation relation.

Although Eqs.~(\ref{eq:tilde-jAt})--(\ref{eq:tilde-rp}) satisfy
the required conditions,
we should not straightforwardly apply them to an actual physical system.
The reason is that an irrelevant contribution is contained in the corrections,
Eq.~(\ref{eq:del-jAt}) and Eqs.~(\ref{eq:del-jAl})--(\ref{eq:del-rp}),
and hence in the final results,
Eqs.~(\ref{eq:tilde-jAt})--(\ref{eq:tilde-rp}), as we show below.
Note that $\varepsilon_{M}/(4\pi v^{2})$ in $\delta\chi_{j A}^{t}$
($\delta\chi_{j A}^{l}$) cancels out the second term of
$\chi_{j A}^{t,{\rm oc}}$ ($\chi_{j A}^{l,{\rm oc}}$),
preserving the gauge invariance.
Let us focus on the remaining terms with $\Lambda(q,\omega)$ in
$\delta\chi_{j A}^{t}$ and $\delta\chi_{j A}^{l}$.
Clearly, they represent the effect of electron excitations
across the cutoff energy.
Since such excitations are artificially allowed as a result of
the modified energy cutoff, the terms with $\Lambda(q,\omega)$
are irrelevant in describing actual situations.
That is, $\delta\chi_{j A}^{t}$ and $\delta\chi_{j A}^{l}$ consist of
the relevant contribution preserving the gauge invariance and
the irrelevant contribution describing the effect of artificial excitations.
$\delta\chi_{\rho A}^{l}$ consists of only the irrelevant contribution.
Note that the irrelevant contributions vanish in the zero-frequency limit
of $\omega \to 0$.
This is not accidental but is guaranteed by
the gauge invariance.~\cite{comment1}

We conclude that only the relevant contributions should be taken into account
in calculating the response functions for a vector potential,
and that the irrelevant contributions vanish in the zero-frequency limit.
Hence, the relevant contributions are selectively picked up
by taking the zero-frequency limit.
This argument justifies the modified energy cutoff procedure
proposed in Sect.~3.

In accordance with the argument given above, we show that the exclusion of
the second terms in Eqs.~(\ref{eq:tilde-jAt})--(\ref{eq:tilde-rp})
is reasonable in actual situations.
Let us focus on the second term of $\tilde{\chi}_{\rho \phi}$.
Owing to its presence, a finite charge density
proportional to $\varepsilon_{M}$ is induced by $\phi$
despite the fact that the valence band is completely filled.
Clearly, this should be regarded as an artifact
induced by the artificial excitations across the cutoff energy.
Hence, the second term should be excluded.
Let us next focus on the second term of $\tilde{\chi}_{j \phi}$.
Since this term is directly related with that of $\tilde{\chi}_{\rho \phi}$
through the charge conservation relation,
it also describes a similar artifact and therefore should be excluded.
The second term of $\tilde{\chi}_{j A}^{l}$ is directly related with
that of $\tilde{\chi}_{j \phi}$ through the gauge invariance.
Furthermore, the second term of $\tilde{\chi}_{j A}^{t}$ must be
identical with that of $\tilde{\chi}_{j A}^{l}$ in the limit of $q \to 0$.
Taking everything into consideration,
we recognize that the second terms describe an artifact
caused by the excitations across the cutoff energy;
thus, they should be excluded.
After the exclusion, $\tilde{\chi}_{j A}^{t}$, $\tilde{\chi}_{j A}^{l}$,
and $\tilde{\chi}_{\rho A}^{l}$ are respectively reduced to
$\chi_{j A}^{t,{\rm mc}}$, $\chi_{j A}^{l,{\rm mc}}$, and $\chi_{\rho A}^{l}$
obtained by using the modified energy cutoff in Sect.~4.
Similarly, $\tilde{\chi}_{j \phi}$ and $\tilde{\chi}_{\rho \phi}$
are respectively reduced to $\chi_{j \phi}$ and $\chi_{\rho \phi}$
given in Sect.~2.

\section{Summary}

We have proposed a modified energy cutoff procedure in terms of
a modified distribution function for electrons
in order to describe the electromagnetic response of Dirac electron systems
in a gauge-invariant manner.
We have shown that the response functions obtained by using this cutoff
satisfy the necessary conditions that are required from the gauge invariance.

Although only the application to a 2D massless Dirac electron system
is presented in this paper,
the modified energy cutoff procedure can be used in various Dirac systems
in any dimension regardless of the presence or absence of a mass gap.
For example, it can be applied to the problem considered
in Ref.~\citen{mizoguchi},
where the superfluid density in a superconducting state
of three-dimensional massive Dirac electrons is calculated
by using a continuum Dirac model under the ordinary cutoff.
The resulting superfluid density does not vanish even in the normal state
without an additional regularization.
If this problem is analyzed by using the modified energy cutoff,
the unphysical contribution [Eq.~(25) of Ref.~\citen{mizoguchi}]
is canceled out by the correction arising
from the intraband term [Eq.~(21) of Ref.~\citen{mizoguchi}].

\section*{Acknowledgment}

This work was supported by JSPS KAKENHI Grant Number JP18K03460.


\begin{thebibliography}{99}
\bibitem{comment0} Since the vector potential is static in this case,
we need not take the zero-frequency limit in the second step.
\bibitem{comment1} Strictly speaking, the gauge invariance guarantees
the vanishing of the irrelevant contribution in the zero-frequency limit
only for a longitudinal vector potential.
In the transverse case,
we may need to employ much more careful treatment in some situations.
\end{thebibliography}
\end{document}